\documentclass[debug]{rmaa}
\usepackage[utf8]{inputenc}
\usepackage{paralist}

\usepackage{psfrag,color}

\title{A Peculiar Galaxy Near M104} 

\author{
  E. Quiroga\altaffilmark{1} 
 }

\altaffiltext{1}{Universidad del Atlantico Medio, Facultad de Comunicacion, Canary Islands, Spain.}
\shortauthor{Quiroga, E.}
\shorttitle{A peculiar galaxy near M104}

\fulladdresses{
\item Lecturer, Universidad del Atlántico Medio, Ctra. de Quilmes, 37, 35017 Tafira Baja, Gran Canaria, Islas Canarias (elio.quiroga@pdi.atlanticomedio.es)}

\abstract{Messier 104, NGC 4594, also known as the Sombrero Galaxy, has been extensively studied, especially its structure and stellar halo. Its abundance of globular clusters has given rise to many theories and much speculation (Ford H. C. et al 1996). But other objects in the vicinity of such a spectacular galaxy are sometimes ignored. While studying HST images available on the HST Legacy website of the halo of M104 (HST proposal 9714, PI: Keith Noll), the author observed at 12:40:07.829 -11:36:47.38 (in j2000) an object about 4 arcseconds in diameter. A study with VO tools suggests that the object is a SBc galaxy with AGN (Seyfert).}

\resumen{Messier 104, NGC 4594, conocida como la Galaxia del Sombrero, ha sido extensamente estudiada, especialmente su estructura y halo estelar. La abundancia de cúmulos globulares que presenta ha dado lugar a muchas teorías y especulaciones (Ford H. C. et al. 1996). Pero otros objetos en las cercanías de una galaxia tan espectacular a veces son pasados por alto. Mientras estudiaba las imagenes del HST disponibles en el sitio web HST Legacy del halo de M104 (propuesta HST 9714, investigador principal: Keith Noll), el autor observa en las coordenadas 12:40:07.829 -11:36:47.38 (en j2000) un objeto de aproximadamente 4 segundos de arco de diámetro. Un estudio con herramientas de VO sugiere que el objeto es una galaxia SBc con un AGN (Seyfert).}

\addkeyword{Virtual observatory tools}
\addkeyword{Catalogs}
\addkeyword{Galaxies: peculiar}
\addkeyword{Galaxies: spiral}
\usepackage{url}
\begin{document}
\maketitle
\section*{METHODS AND DISCUSSION}
\label{sec:intro}

The author, studying Hubble Space Telescope (HST) images of M104 available on the HST Legacy website (HST proposals 9714, PI: K. Noll, and 13364, PI: D. Calzetti), observes at coordinates 12:40:07.829 -11:36:47.38 an object located in the halo of the galaxy, about 4 arcseconds in diameter. It is catalogued by Simbad as a Globular Cluster Candidate. The Nasa/IPAC Extragalactic Database (NED)\footnote{\url{https://ned.ipac.caltech.edu/byname?objname=SSTSL2+J124007.83-113647.1&hconst=67.8&omegam=0.308&omegav=0.692&wmap=4&corr_z=1}} shows its classification as an IrS (Infrared Source), not as a galaxy, with an available spectral energy distribution (SED) plot. A search of the Pan-STARRS1 data archive returns objName PSO J190.0326-11.6132, which in VizieR shows a more complete SED plot \footnote{\url{http://vizier.cds.unistra.fr/vizier/sed/?-c=12+40+07.83277+-11+36+47.0442&c.rs=5.0}} with data from 5 catalogues: PAN-STARRS PS1, SDSSz, 2MASSJ, Spitzer:IRAC3.6 and VISTA:Z. After a new search in NED, the object in region \#710, SSTSL2 J124007.83-113647.1, seems to be the one.

The author has compared the object's SED plot with those of spiral galaxy models in Optical-NIR-MIR (Nicole P. V. 2012), and the object's NED plot points fit quite well. 

HST Legacy has some images taken with 435 and 555 filters, which are equivalent to B, V and R filters. So the bandwidth is from B to NIR. The results of an RGB of the images in Aladin Sky Atlas, (Figure 1b) show (limited by the resolution) a galactic centre and a central bar with reddish hues (555 filter, R of the Wide Field and Planetary Camera 2 [WFPC2]) and a blue ring around it (435 filter, B+V of the Advanced Camera for Surveys [ACS]). This may suggest a central region of cool, old stars and two possible spiral arms with star-forming regions.

\begin{figure}
    \centering
    \includegraphics[width=1\linewidth]{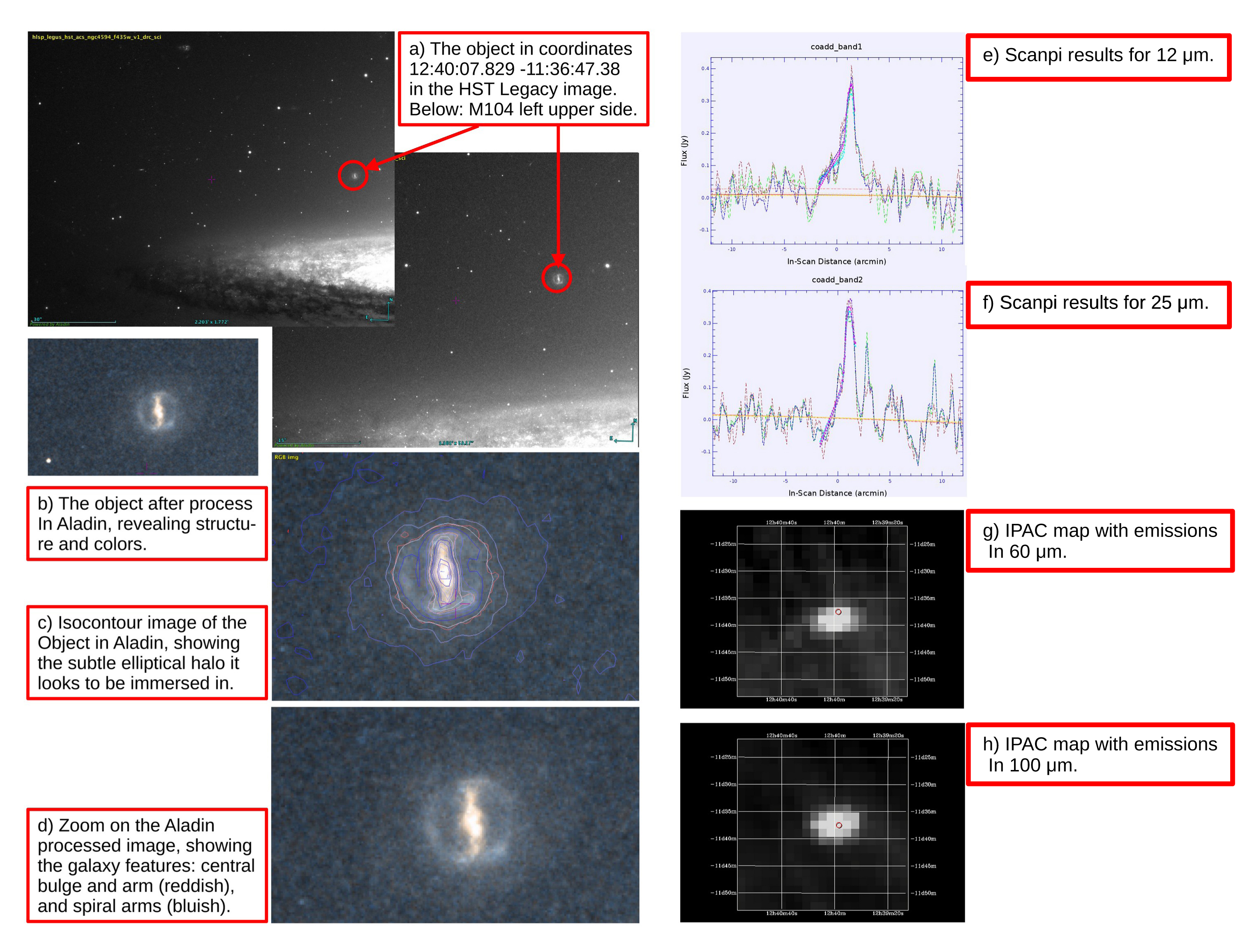}
    \caption{ The adjusted and zoomed RGB of the object in Simbad (a, b, c), the Scanpi results for 12 micrometers (d) and for 25 micrometers (e), and the IPAC map with emissions at 60 (f) and 100 micrometers (h) in the halo area shown in b.}
    \label{fig:enter-label}
\end{figure}

In addition, the approximately circular object (which appears to be face-on) has a subtle elliptical halo twice its size (Figure 1c), possibly a feature of the PSF due to the low signal-to-noise ratio in the area. The author obtained an IPAC map (with data from the Spitzer Space Telescope, Wide-field Infrared Survey Explorer WISE) showing emissions at 60 and 100 micrometers, within an approximate radius of 7 arcseconds around the coordinates of the object (Figures 1g and 1h), which is 4 arcseconds wide; it shows that the halo has an infrared emission, which is not a feature of the PSF. In Figure 1e, Scanpi results for the object show a flux gap at 12 micrometers (silicate absorption feature), which may be indicative of an AGN Seyfert galaxy (Koehler \& Li 2010), (Hao et al 2007). Also shown in Figure 1f are the Scanpi results at 25 micrometers, which show a peak. The 25 micrometers emission is a good measure of AGN luminosity (Severgnini, P. et al 2012). The WISE satellite offers a spatial resolution of approximately 6 arcsec \footnote{\url{https://wise2.ipac.caltech.edu/docs/release/allsky/}} while Spitzer provides about 2 arcsec \footnote{https://lweb.cfa.harvard.edu/\texttt{\~{}}mmarengo/me/irac.html},  so it is probable that the origin of the infrared emission is the galaxy, not the halo. 

The silicate absorption feature may indicate that this galaxy once hosted massive stars (type O or B) capable of burning silicon in their final stages (Vollmer et al 2008). Supernova explosions may also have created a silicon-rich interstellar medium around the galaxy's core (Woosley, 2005).

In terms of redshift and angular size, Simbad\footnote{http://simbad.cds.unistra.fr/simbad/sim-id?Ident=\%4012370942\&Name=PSO\%20J190.0301-11.6124\&submit=submit} gives us a radial velocity for the object of 1359 km/s and a z = 0.004545 +/- 0.000027. M104 has a z = 0.003416, so these data should be taken with the necessary caution (the measured redshift of a galaxy results from a combination of the cosmological redshift due to the expansion of the Universe and a Doppler shift due to the peculiar motions of the galaxy, and it is widely accepted that for galaxy collections, if the difference in redshift corresponds to a velocity difference below 500 km/s, or 1000 km/s for clusters, one cannot affirm with certainty that the objects are gravitationally unbound (Capelato, H.. V. et al 1991)). In this case the velocity difference is about 300 km/s. The author cannot confirm that the object is not gravitationally bound to M104, so it could be a satellite of the Sombrero galaxy with an angular size around 0.3 kpc.

As an alternative scenario, assuming the object is not associated with M104, it could be located at 20 Mpc $ (H0=67.04, \Omega m=0.3183  $ and $ \Omega \Lambda =0.6817)$ with an angular size $ \approx 22 Kpc $, i.e. $ \frac{2}{3} $ the size of the Milky Way. The object has a Fe/H metallicity of -0.309 dex (Alves-Brito, A. et al 2011), a relatively low value (metal-poor).
Finally, there is an entry in HEASARC \footnote{bit.ly/46x1yrJ} from the CXOGSGSRC table of the CHANDRA observatory database for the object at coordinates 12 40 06.24 -11 36 47.7 with an X-ray flux emission of $ 2.410e-15 erg/cm^{2}/s $, i.e. an X-ray luminosity of $ L_X \approx 1.80 \times 10^{43} erg/s $, assuming the alternative scenario with a distance of 20 Mpc; such a luminosity may indicate an active galactic nucleus (AGN), which is in the range observed for Seyfert galaxies (Panessa, F. et al, 2006). Further studies would be needed to determine whether this is a Type 1 or Type 2 AGN.

\section*{CONCLUSIONS}

The author has recategorised an object erroneously catalogued by Simbad as a Globular Cluster Candidate as an SBc AGN Seyfert galaxy, z = 0.004545 +/- 0.000027, using VO tools. The Aladin Sky Atlas RGB suggests an SBc galaxy with a dominant central arm, nucleus and possibly two spiral arms with hot young stars and dust. VizieR and NED SED plots are consistent with spiral galaxies. The object has a subtle elliptical halo twice the size of the central object, almost unnoticeable in the visible, with far-infrared emission in a radius of 7 arcseconds. Scanpi results at 12 micrometers  (silicate absorption feature) and a peak at 25 micrometers suggest a possible AGN. The object has an X-ray emission luminosity of $ L_X \approx 1.80 \times 10^{43} erg/s $ assuming a distance of 20 Mpc, suggesting a possible Seyfert galaxy in the AGN Unification Model (Singh, V. et al., 2011).

The author recommends adding the object to the "Galaxy" category, especially in public catalogues such as NED or Simbad, where it is still classified as an "Infrared Source" or "Globular Cluster Candidate". The author proposes that it be named The Iris Galaxy.

\section*{ACKNOWLEDGEMENTS}

Thanks to Adrian B. Lucy in the Space Telescope Science Institute for his kind help. 

Thanks to Alex Fraser for the English revision.

Thanks to Agustin Trujillo (ULPGC) for the LaTeX translation help.

This research has made use of the Aladin Sky Atlas developed at CDS, Strasbourg Observatory, France,  the NASA/IPAC Extragalactic Database (NED), the VizieR catalogue access tool, the Pan-STARRS1 Surveys (PS1), the Two Micron All Sky Survey (2MASS), the SDSS catalogue, the SIMBAD astronomical database, the European Space Agency (ESA) Gaia mission, and the Chandra X-Ray Center (CXC), operated for NASA by the Smithsonian Astrophysical Observatory. This research is based on observations made with the NASA/ESA Hubble Space Telescope obtained from the Space Telescope Science Institute.

This research has made use of the NASA/IPAC Infrared Science Archive, which is funded by the National Aeronautics and Space Administration and operated by the California Institute of Technology.

\section*{DATA AVAILABILITY STATEMENT}

VO tables for the VizieR and NED SEDs plus data in Simbad and Heasarc are in the footnote links.

The short link for HEASARC stays for the url: \url{ https://heasarc.gsfc.nasa.gov/db-perl/W3Browse/w3hdprods.pl?files=P&Target=heasarc%5Fxray%20%7C%7C%7C%5F%5Frow%3D1386796%7C%7C&Coordinates=Equatorial&Equinox=2000.}

\section*{REFERENCES}
\hangindent=1cm
Alves-Brito, A., Hau, G. K., Forbes, D. A., Spitler, L. R., Strader, J., Brodie, J. P., \& Rhode, K. L. 2011. Monthly Notices of the Royal Astronomical Society, 417(3), 1823-1838. https://doi.org/10.48550/arXiv.1107.0757

\hangindent=1cm
Capelato, H. V., Mazure, A., Proust, D., Vanderriest, C., Lemonnier, J. P., \& Sodre Jr, L. 1991. Astronomy I\& Astrophysics Supplement Series, 90, 355-364. https://adsabs.harvard.edu/full/1991A

\hangindent=1cm
Ford H. C. et al. 1996. The Astrophysical Journal , 458: 455-466, 1996 20 February. 

\hspace{0.4cm} 
https://ui.adsabs.harvard.edu/link\_gateway/1996ApJ...458..455F

\hspace{0.4cm} 
/doi:10.1086/176828

\hangindent=1cm
Koehler, M., \& Li, A. 2010. Monthly Notices of the Royal Astronomical Society: Letters, 406(1), L6-L10. https://doi.org/10.48550/arXiv.1210.6562

\hangindent=1cm
Noll, K. 2002. HST Proposal, 9714.

\hangindent=1cm
Panessa, F., Bassani, L., Cappi, M., Dadina, M., Barcons, X., Carrera, F. J., Ho, L.C. \& Iwasawa, K. 2006. Astronomy I\& Astrophysics, 455(1), 173-185. https://doi.org/10.1051/0004-6361:20064894

\hangindent=1cm
Severgnini, P., Caccianiga, A., I\& Della Ceca, R. 2012. Astronomy I\& Astrophysics, 542, A46. https://doi.org/10.48550/arXiv.1204.4359

\hangindent=1cm
Singh, V., Shastri, P., \& Risaliti, G. 2011. Astronomy I\& Astrophysics, 532, A84. https://doi.org/10.48550/arXiv.1101.0252

\hangindent=1cm
Spoon, H. W., Marshall, J. A., Houck, J. R., Elitzur, M., Hao, L., Armus, L., ... \& Charmandaris, V. 2006. The Astrophysical Journal, 654(1), L49. https://iopscience.iop.org/article/10.1086/511268

\hangindent=1cm
Vollmer, C., Hoppe, P., \& Brenker, F. E. 2008. The Astrophysical Journal, 684(1), 611. https://iopscience.iop.org/article/10.1086/589913

\hangindent=1cm
Woosley, S., \& Janka, T. 2005. Nature Physics, 1(3), 147-154. 

\hspace{0.4cm} 
https://doi.org/10.48550/arXiv.astro-ph/0601261

\vfill Elio Quiroga Rodríguez. Lecturer, Universidad del Atlántico Medio, Facultad de Comunicación, Ctra. de Quilmes, 37, 35017 Tafira Baja, Gran Canaria, Islas Canarias (elio.quiroga@pdi.atlanticomedio.es)

\end{document}